\documentclass[namedreferences]{SolarPhysics}
\usepackage[optionalrh]{spr-sola-addons} 

\usepackage{graphicx,times,url,color}

\newcommand{\etal}{{\it et al.}}
\newcommand{\eg}{{\it e.g.}}
\newcommand{\degr}{\mbox{$^\circ$}}%

\newcommand{\aap}{    {\it Astron. Astrophys.}}

\newcommand{\an}{     {\it Astron. Nachr.}}
\newcommand{\apj}{    {\it Astrophys. J.}}

\newcommand{\mnras}{  {\it Mon. Not. Roy. Astron. Soc.}}
\newcommand{\nat}{    {\it Nature}}

\newcommand{\solphys}{{\it Solar Phys.}}
\newcommand{\ssr}{    {\it Space Sci. Rev.}}

\begin{document}
\begin{article}
\begin{opening}

\title{Towards Waveform Heliotomography: Observing Interactions of 
       Helioseismic Waves with a Sunspot}
\author{Junwei~Zhao \sep
        Alexander~G.~Kosovichev \sep
        Stathis~Ilonidis }
\runningauthor{Zhao \etal}
\runningtitle{Interaction of Waves with a Sunspot}
\institute{W.W.Hansen Experimental Physics Laboratory, Stanford University,
           Stanford, CA94305-4085, USA
           email: \url{junwei@sun.stanford.edu} \\
          }

\begin{abstract}
We investigate how helioseismic waves that originate from effective point 
sources interact with a sunspot. These waves are reconstructed from 
observed stochastic wavefields on the Sun by cross-correlating photospheric 
Doppler-velocity signals. We select the wave sources at different locations 
relative to the sunspot, and investigate the $p$- and $f$-mode waves 
separately. The results reveal a complicated picture of waveform perturbations
caused by the wave interaction with the sunspot. In particular, it is 
found that for waves originating from outside of the sunspot, $p$-mode
waves travel across the sunspot with a small amplitude reduction and 
slightly higher speed, and wave amplitude and phase get mostly restored
to the quiet-Sun values after passing the sunspot. The $f$-mode wave
experiences some amplitude reduction passing through the sunspot, and 
the reduced amplitude is not recovered after that. The wave-propagation 
speed does not change before encountering the sunspot and 
inside the sunspot, but the wavefront becomes faster than the reference
wave after passing 
through the sunspot. For waves originating from inside the sunspot 
umbra, both $f$- and $p$-mode waves show significant amplitude reductions 
and faster speed during all courses of propagation. A comparison of positive 
and negative time lags of cross-correlation functions shows an apparent 
asymmetry in the waveform changes for both the $f$- and $p$-mode waves. 
We suggest that the waveform variations of the helioseismic waves 
interacting with a sunspot found in this article can be used for developing
a method of waveform heliotomography, similar to the waveform tomography 
of the Earth.
\end{abstract}


\end{opening}

\section{Introduction}
  \label{S1}
Local helioseismology techniques, including time-distance helioseismology
\cite{duv93}, acoustic imaging \cite{cha97}, and helioseismic 
holography \cite{lin97}, have been widely used to study solar acoustic 
wave travel times or phase shifts, which are then used to invert 
interior structures and flow fields beneath sunspots (\eg{} \opencite{kos00}; 
\opencite{giz00}; \opencite{zha01}; \opencite{sun02}; \opencite{zha10}). 
These inversions treated the travel-time shifts as caused by subsurface 
sound-speed perturbations and plasma flows, and they have provided us the 
tomographic images of the sunspots' interior properties. However, a 
picture of how helioseismic waves originating from a point source, including 
both surface gravity ($f$-mode) waves and acoustic ($p$-mode) waves, 
interact with sunspots has not been systematically studied and visualized
observationally.  This is complicated because solar 
waves are excited stochastically (except for rare flare events such as 
reported by \opencite{kos98}), and thus the observed wave field represents 
a superposition of waves excited by many sources randomly distributed in 
space and time.

To understand the effects of the magnetic field on the propagation of 
helioseismic waves, theoretical efforts have been made to investigate MHD mode
conversions, wave absorptions, and wave interactions with inclined magnetic
field lines \cite{cal05, cal08, sch08}. Numerical simulations have 
also provided useful approaches to the understanding of
how helioseismic waves travel through magnetized areas in more 
realistic conditions. Recently, \inlinecite{cam08} studied how $f$-mode 
waves interact with magnetic fields by employing an MHD simulation code 
\cite{cam07}, and then they compared their simulation results with the observed 
signals in the photosphere by summing cross-correlations along a 
straight line. A similar approach was taken to study $p$-mode interactions
with a sunspot more recently \cite{cam10}. This has given a successful
example of how the solar interior properties can be studied by forward
modeling. MHD simulations of helioseismic waves, both $f$- 
and $p$-mode, have also been carried out to study various problems 
of wave excitation and wave -- magnetic-field interactions in 3D 
models by numerous authors, \eg{} \inlinecite{kho09}, \inlinecite{par09},
\inlinecite{she09}, and \inlinecite{ste10}. Some of these theoretical 
models and simulations consider waves from a point source. Thus, 
for comparisons with observations, it is necessary to extract from the 
observational data the wave propagation and wave -- sunspot interaction
corresponding to a point source from the observational data.

In this article, we employ a time-distance cross-correlation approach to 
obtain the wave signals (Green's function) from localized point sources, and
give pictures of how the $f$- and $p$-mode waves interact with a sunspot. 
We describe our data and method in Section~\ref{S2}, and 
present results in Section~\ref{S3}. In Section~\ref{S4}, we present
results obtained for negative time lags of the cross-correlation
functions (``incoming'' waves). Summary and Discussion follow in 
Section~\ref{S5}.

\begin{figure}
\centerline{\includegraphics[width=1.0\textwidth]{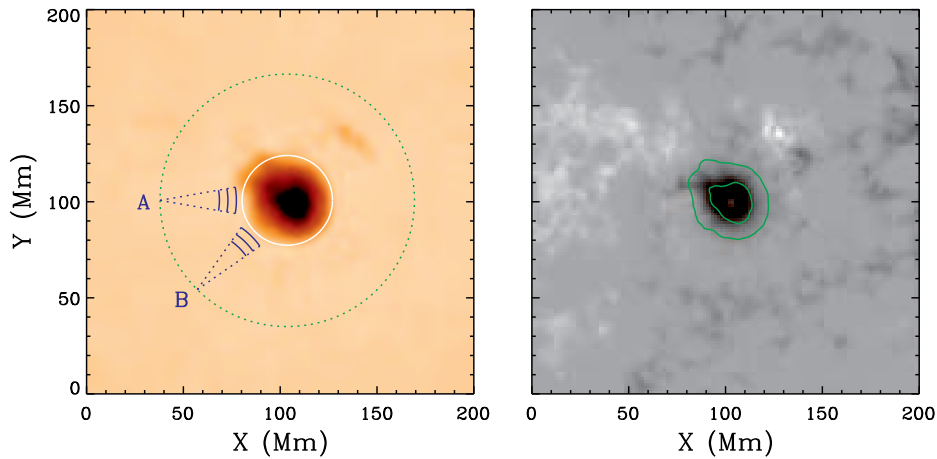} }
\caption{{\it Left}: Schematic plot over the intensity map of the active
region showing how cross-correlation functions are obtained and averaged. 
Points ``A" and ``B" are two examples of wave sources, and the green dotted 
circle shows where wave sources are selected. The white circle indicates
that this sunspot is not far from being axisymmetric. {\it Right}: 
Magnetogram of this active region, averaged over the entire observation
period. The red contours represent boundaries of the sunspot umbra and 
penumbra. }
\label{schematic}
\end{figure}

\section{Data and Method}
  \label{S2}
For this analysis, we select a stable, long-lived active region, NOAA AR9787, 
observed by the Solar and Heliospheric Observatory/{\it Michelson Doppler 
Imager} (SOHO/{\it MDI}; \opencite{sch95}) in full-disk resolution Doppler 
velocities. This same active region was also studied by \inlinecite{cam08},
\inlinecite{cam10}, and other works as summarized in \inlinecite{giz09}.
Our analysis period covers from 00:00UT 22 January through 
23:59UT 26 January, 2002, a total of five entire days. The shape of the
sunspot inside this active region did not significantly change during 
the observational period. For convenience, the entire data sequence 
is divided into ten twelve-hour segments, and each segment is analyzed 
separately. The cross-correlation functions are obtained from each
segment, and then these functions are averaged to get the final waveforms. 

As shown in Figure~\ref{schematic}, the sunspot located in this active
region is not perfectly round but close enough to be considered 
axisymmetric. Theoretically, if we select location ``A" in 
Figure~\ref{schematic} as a wave source, we can reconstruct the wavefield 
originating from ``A" as a function of time and two-dimensional
space by cross-correlating solar oscillation signals at ``A" with signals
at all other locations inside the area of interest \cite{cla92}, using
$$\psi(\tau,\mathbf{d}) = \int f_\mathrm{A}(t) f(t+\tau, \mathbf{d}) 
\mathrm{d}t, $$ where $f_\mathrm{A}(t)$ is the observed oscillation signal
at location ``A", $f(t, \mathbf{d})$ is the oscillation signal observed 
at a location $\mathbf{d}$ relative to ``A", and $\tau$ is the time lag
between the signals. \inlinecite{zha07} have used a similar approach 
to reconstruct acoustic wave propagation (wavefronts from a point source)
in the photosphere and in the interior using the realistic numerical 
simulations of solar convection \cite{ben06}. The cross-correlations provide 
the wave signals (or Green's function) from an effective point 
source located at ``A". However, to obtain a good signal-to-noise ratio,
some averaging over multiple locations is necessary. If we assume that 
this sunspot is axisymmetric, then we can rotate the wavefield obtained
by cross-correlating signals at ``B" with all other locations in the
area of interst by a certain angle, so that points ``A" and ``B" coincide, and 
then we average these two wavefields. Similarly, all locations with the same
distance to the sunspot's center, shown as green dotted circle,
can be used for averaging. The averaged wavefield provides a Green's 
function of helioseismic waves propagating from an effective point source
to all other areas, including the sunspot.
In practice, we do additional averaging. When the wave source is located 
outside the sunspot's umbra, we select an area of $3\times3$ pixels as the 
center of the dotted green circle, and compute the wavefields for each of 
these pixels separately. Then we average these wavefields to increase the
signal-to-noise ratio. 

However, when the wave source is located inside the sunspot 
umbra, we use all points inside the sunspot umbra for the mean wavefield 
calculations, and in this case no wavefield rotations are performed before 
the averaging. It is clear that
different number of pixels is used when ``A" is located at different 
distances from the sunspot center, and this would introduce different 
signal-to-noise ratios in our averaged cross-correlation functions. 
We compare cross-correlation functions from the sunspot
with those from the quiet Sun following the same procedure to reduce
this effect, although it is not quite clear how the noise influences
our final results.

Our method is similar to the method by \inlinecite{cam08} in retrieving
wave propagation by calculating cross-correlations. However, we average
our signals using a different geometry such that our resultant wave 
propagation is from an effective point source. This method makes the 
study of waves' interaction with sunspots simple and straightforward. It
also provides observational data that can be easily compared with numerical 
simulations of waves from point sources. Additionally, the point-source
signals effectively represent a Green's function of the wave, which
is a primary tool in waveform tomography \cite{sch09}. 

\begin{figure}
\centerline{\includegraphics[width=1.0\textwidth]{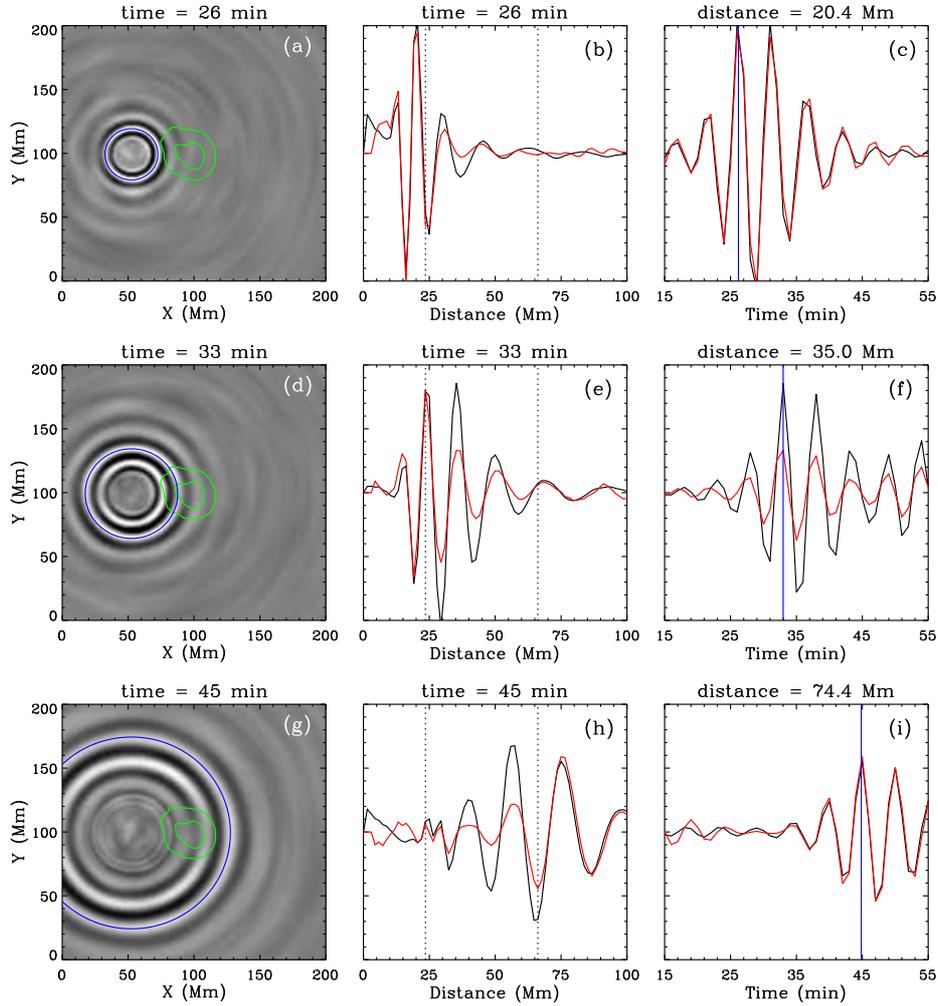} }
\caption{Selected snapshots showing the interaction of $p$-mode waves with
the sunspot when the wave source is located outside the sunspot. Left column 
shows two-dimensional images of the waveform. Red contours indicate 
the sunspot umbra and penumbra boundaries, and the blue circles show
the travel distances obtained for these time moments from the time--distance 
relation of the quiet Sun. In the middle column, red curves show the 
waveforms along the line connecting the wave source and the sunspot 
center, i.e. $0\degr$ angle line, in the left images.
Black curves are the corresponding waveforms obtained from 
the quiet Sun, and these are used as reference waves. Vertical dotted 
lines indicate the two boundaries of the sunspot. In the right column, 
red curves show oscillations at the location where the blue 
circles (in the left column) meet the $0\degr$ angle line, and dark 
curves are oscillations from the reference wave at the same
propagation distance. The vertical blue lines indicate the reference 
acoustic phase travel times of the quiet Sun. The travel times are
obtained by fitting oscillation functions using Gabor wavelet function. 
Vertical scales for the middle and right columns are arbitrary. }
\label{p_1}
\end{figure}

\begin{figure}
\centerline{\includegraphics[width=1.0\textwidth]{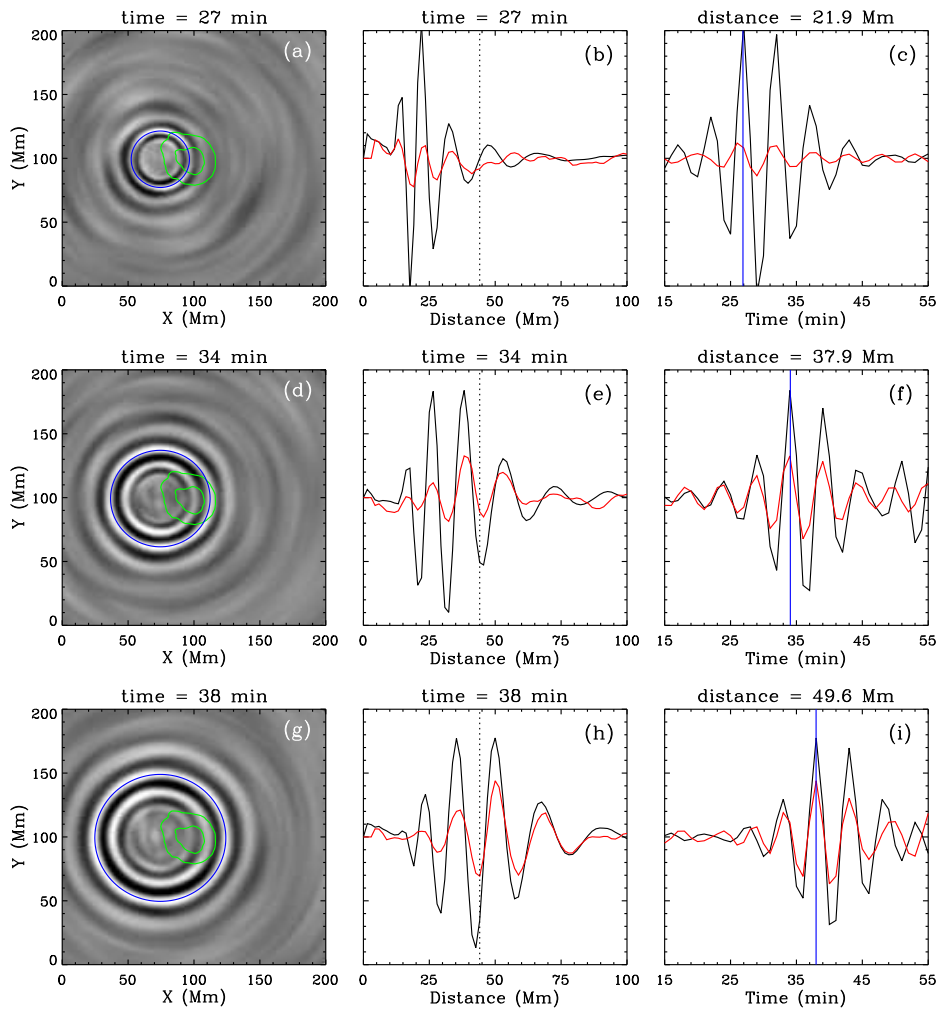} }
\caption{Same as Figure~\ref{p_1}, but for a wave source located
at the boundary of the sunspot penumbra. The dotted lines in the middle column
indicate the penumbra boundary on the opposite side of the wave source. }
\label{p_2}
\end{figure}

\begin{figure}
\centerline{\includegraphics[width=1.0\textwidth]{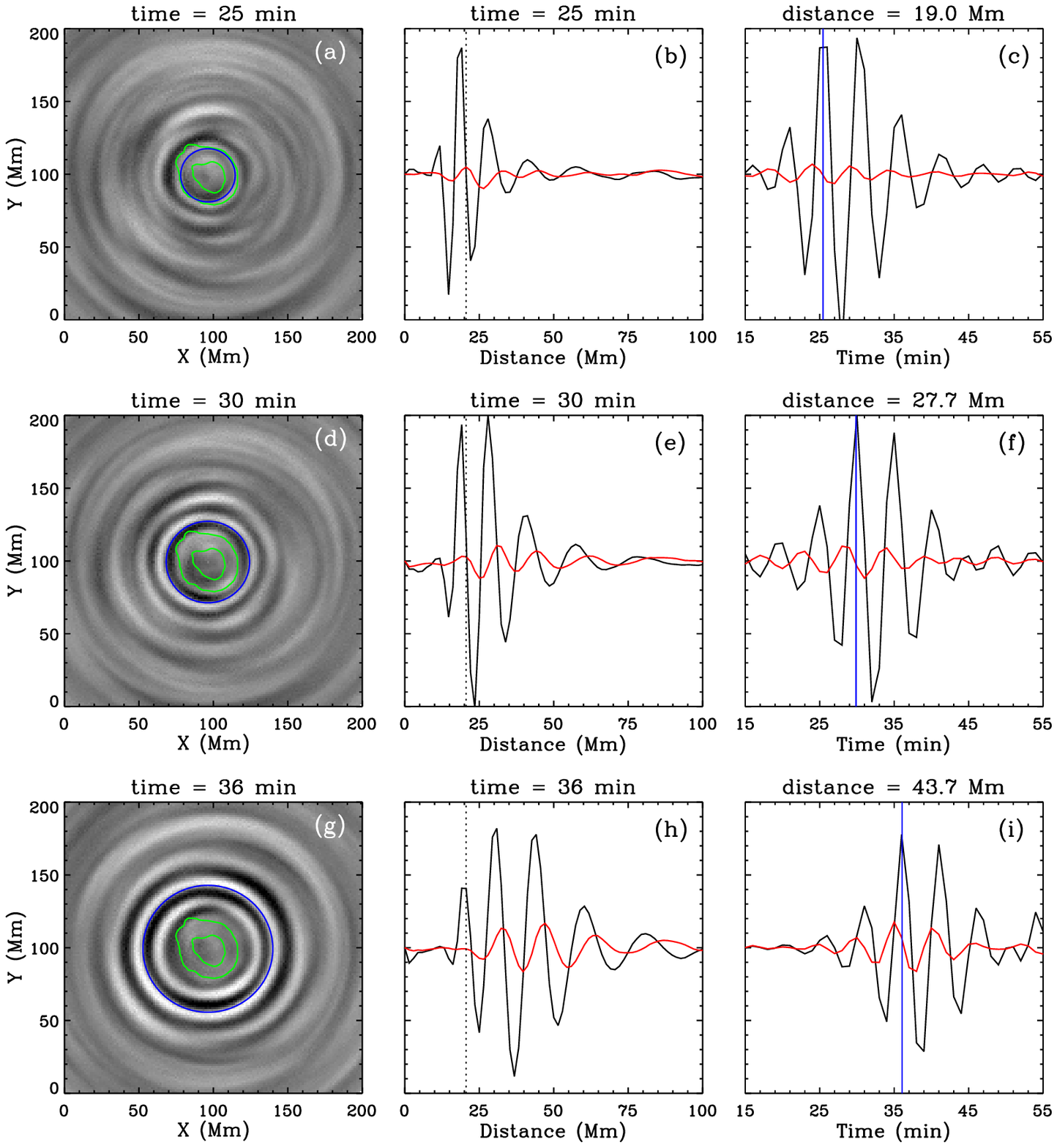} }
\caption{Same as Figure~\ref{p_1}, but when the wave sources are located
inside the sunspot umbra and averaged over the whole umbra. The dotted 
line in the middle column indicates the sunspot boundary. Red curves in 
the middle and right columns are obtained by averaging all propagation 
directions.}
\label{p_3}
\end{figure}

\section{Results}
  \label{S3}
Because the solar $f$- and $p$-mode waves have very different properties, we 
first separate the $f$- and $p$-mode signals in the Fourier domain, and then
analyze these two types of waves separately. No other filters, such as a
phase-speed filter \cite{kos00} or a ridge filter \cite{bra08}, are applied 
after the mode separation. We use frequencies above 1.8 mHz for the $f$-mode,
and all frequencies above the $f$-mode ridge for the $p$-mode analysis.

\subsection{Results for $p$-Mode Waves}
  \label{S3_1}
For the $p$-mode waves, we study three different cases: when the wave source 
is located outside the sunspot, at the boundary of the sunspot penumbra 
and the quiet Sun, and inside the sunspot umbra.

Figure~\ref{p_1} shows results of the first case, at the 
26th, 33rd, and 45th minute of the wave propagation, when the wave 
source is located 44 Mm from the sunspot center. The selection of 
this distance is arbitrary, and all pixels at the same distance are used 
to calculate the cross-correlations. In Figure~\ref{p_1}, 
the left column shows two-dimensional wave propagation images, the 
middle column shows comparisons of the waveforms of the waves traveling
through the sunspot and in the quiet Sun, and the right column shows 
comparisons of the corresponding oscillations at the given
distances. The comparisons with the quiet Sun data in the middle and 
right columns give a better visualization of differences in the wave 
propagation and acoustic travel times for the waves with and without
interaction with the sunspot. For this case and all following cases,
the reference waveforms are obtained for the quiet Sun following 
exactly the same procedures as for the sunspot study so as to avoid issues 
such as normalization or different signal-to-noise ratios. 
At $\tau = 26$ minute, the wavefront enters 
the sunspot, and the strongest wave peak is right outside the boundary 
of the sunspot. Figure~\ref{p_1}(b) shows more clearly that inside the 
sunspot boundary the wavefront experiences some amplitude reduction and 
also moves slightly ahead of the reference wave that is obtained from 
the quiet Sun. The oscillation, shown in Figure~\ref{p_1}(c), 
obtained at the strongest wave peak that is still located outside the 
sunspot, does not show a noticeable travel-time shift relative to 
the reference oscillation. At $\tau = 33$ minute, 
the major wave peak reaches the sunspot umbra, and this wave peak 
experiences a significant amplitude reduction and moves slightly ahead 
of the reference wave, as can be seen in Figure~\ref{p_1}(e). The 
oscillations from this wave peak shows that the travel time is 
approximately 0.3 minutes faster than the reference wave.
At $\tau = 45$ minute, the wavefront passes through the sunspot area and 
arrives at the other side. The comparison of both waveforms and oscillations
demonstrates that once the $p$-mode waves arrive to the other 
side of the sunspot, the wave does not have a noticeable amplitude 
reduction or travel-time shift. At first sight, it may look strange 
that the acoustic waves return to their original properties after passing 
through the sunspot. However, one should keep in mind that the acoustic 
waves travel through the interior, and that their traveling depths 
depend on travel distances. The waves that appear on the surface at 
larger distance travel through the deeper interior, and appear relatively
unaffected by the sunspot. 

Figure~\ref{p_2} shows the results when the wave source is located at the 
boundary of the sunspot penumbra and the quiet Sun. At $\tau = 27$ minute,
the strongest wave peak reaches the central sunspot area. The comparisons 
with the reference wave, shown in Figure~\ref{p_2}(b) and (c), demonstrate 
that the wave experiences substantial amplitude reduction, and that this 
wave peak moves approximately 0.5 minutes ahead of the reference wave. At 
$\tau = 34$ minute, the major peak is still inside the sunspot and shows 
faster propagation, while the wavefront outside of the sunspot shows 
slightly slower propagation. Oscillation sampled 
right inside the sunspot boundary, shows a slight negative travel 
time shift of approximately 0.3 minutes. When the large fraction of the 
wave passes through the sunspot area at $\tau = 38$ minute, shown
in Figure~\ref{p_2} (d) -- (e), it seems that the wave has roughly the same 
propagation time as the reference quiet-Sun wave, and experiences some 
amplitude reduction compared with the reference wave profile. However, 
this amplitude reduction may be due to the acoustic emissivity reduction 
at the wave source \cite{cho09, ilo10}. 

Figure~\ref{p_3} shows the results when the wave source is located inside
the sunspot umbra. The red curves in the middle and right columns of
the figure are obtained by averaging all propagation directions to 
enhance the signal-to-noise ratio. The reference curves are obtained by 
the same means. It is clear from the figure that at different propagation 
time, the wave experiences substantial amplitude reduction, and negative 
travel-time shifts of approximately 1.0 minute for nearly all locations. 
This is true for the wavefronts that are still inside the sunspot, and 
is also true when the wave is very far from the wave source, close to the 
edge of the analysis area. It is also clear from Figures~\ref{p_3}(c), (f), 
and (i) that the waves originating from the sunspot umbra have roughly 
the same oscillation period as waves originating from the quiet Sun, 
i.e. the travel-time shifts are roughly the same for different oscillation 
peaks.  

\begin{figure}
\centerline{\includegraphics[width=0.78\textwidth]{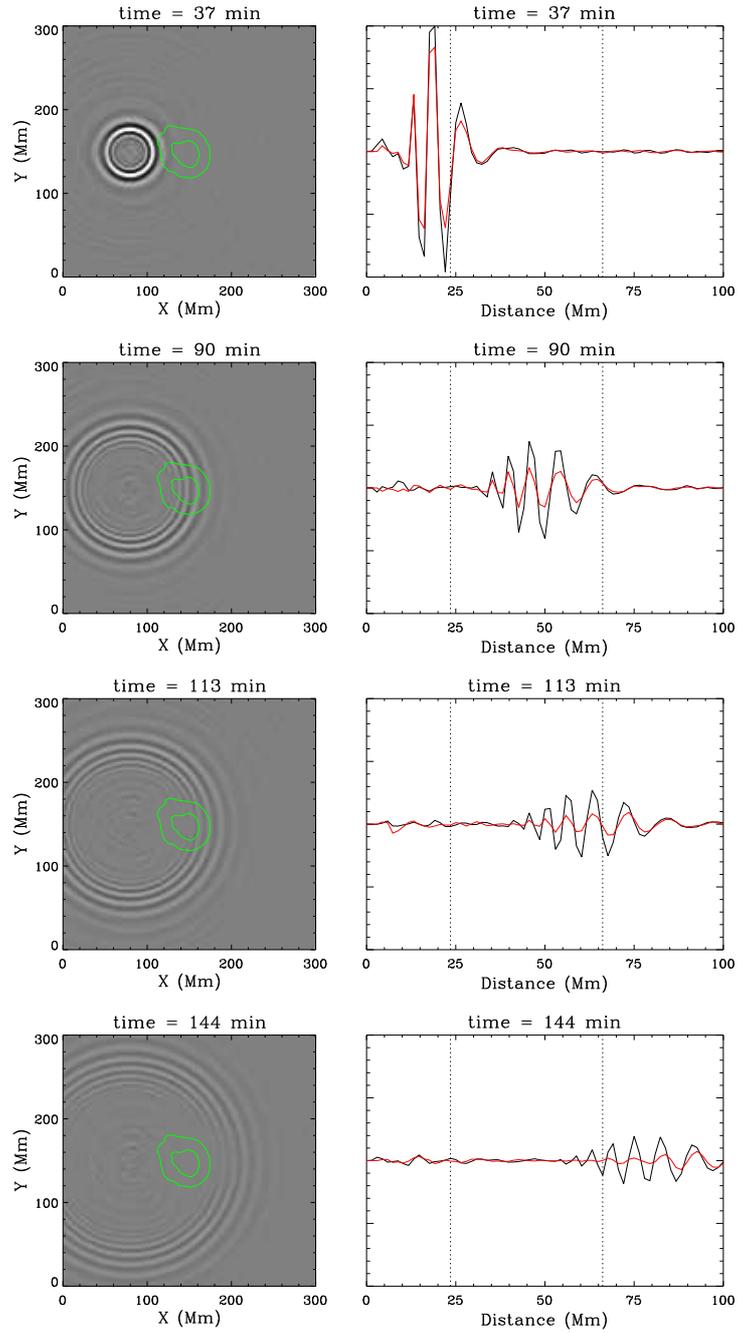} }
\caption{Selected snapshots showing the $f$-mode wave -- sunspot interaction
when the wave source is located outside the sunspot. Left column shows 
two-dimensional images of wave propagation, with red contours indicating 
the sunspot umbra and penumbra boundaries. In the right column, red curves
are horizontal cut through the line linking the wave source and the sunspot
center, i.e. $0\degr$ angle. Black curves are averaged from the quiet 
Sun reference wave. Vertical dotted lines represent the sunspot boundaries. }
\label{f_1}
\end{figure}

\begin{figure}
\centerline{\includegraphics[width=0.65\textwidth]{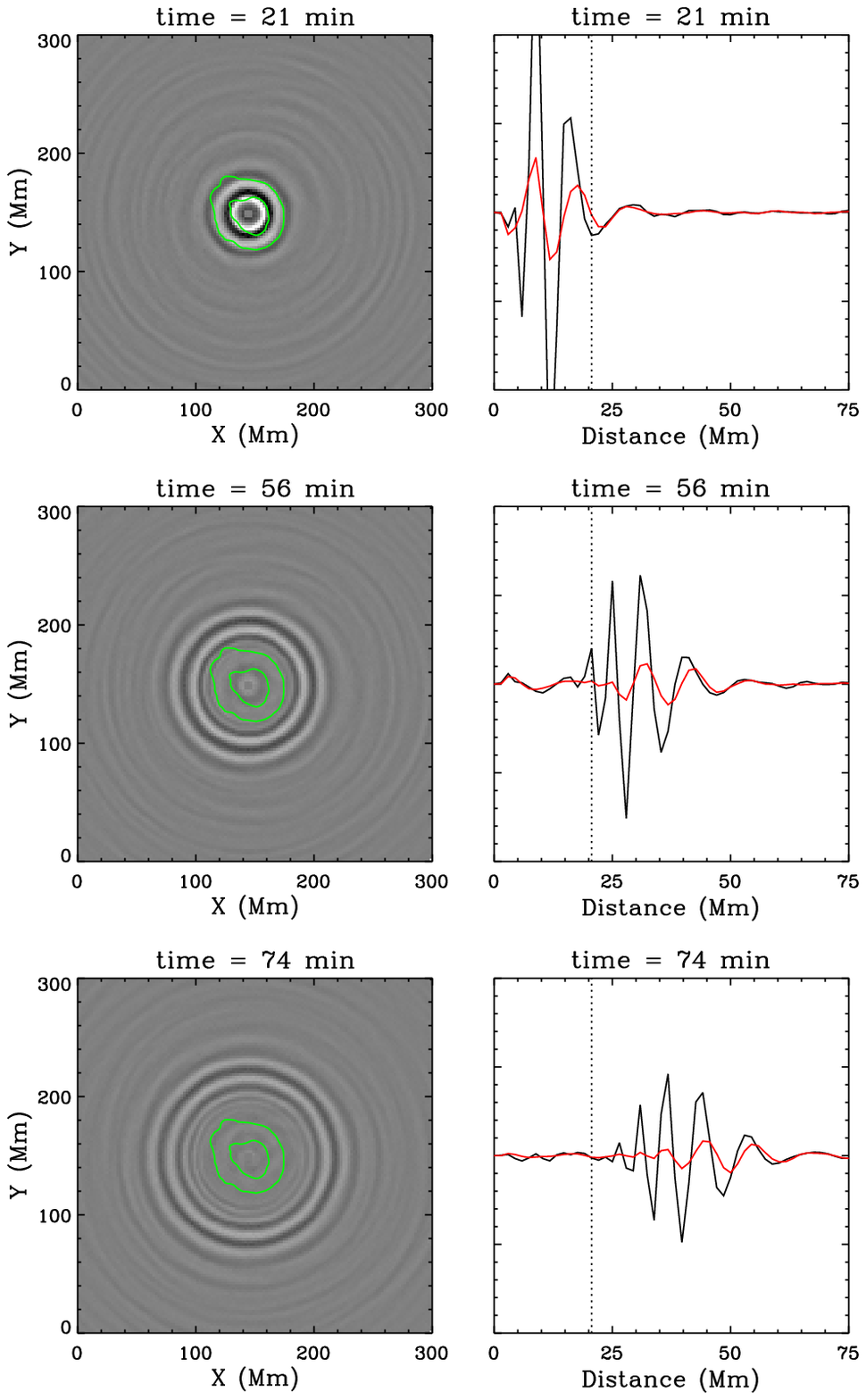} }
\caption{Same as Figure~\ref{f_1}, but when the wave source is located 
inside the sunspot umbra. Dotted lines in the right column represent 
the sunspot boundary. Red curves are obtained by averaging all wave 
propagation directions. }
\label{f_2}
\end{figure}

\subsection{Results for $f$-mode waves}
  \label{S3_2}
For the $f$-mode waves, we study two different cases: when the wave source
is located outside the sunspot and inside the sunspot umbra.

Figure~\ref{f_1} shows selected snapshots when the wave source 
is located outside the sunspot, 44 Mm from the sunspot center.
Clearly, the $f$-mode wave propagates much slower than the $p$-mode waves. 
It can be found that the wave amplitude reduction starts soon after the
wave encounters the sunspot boundary, and the reduction becomes more 
significant when the wave enters and passes through the sunspot.
Unlike the $p$-mode case, the amplitude reduction does not get recovered 
after the wave passes the sunspot. For the propagation
speed, we find that before the wave encounters the sunspot
boundary and during its propagation inside the sunspot (two sample
snapshots at $\tau = 37$ and $90$ minute are shown in Figure~\ref{f_1}),
there is no clear difference in the speed or travel times from
the quiet-Sun reference. However, when the wavefront passes the sunspot 
boundary to the outside, at $\tau = 113$ minute, 
the first peak of the wavefront is clearly ahead of the reference, 
while the third and fourth peak (counting from the right hand side) 
are slightly behind the reference wave. This probably indicates a change
of the oscillation frequencies and/or the wave dispersion relation inside 
the magnetized areas. At $\tau = 144$ minute when the wave nearly 
completely reaches the other side of  the sunspot, 
the first two wave peaks on the right hand side are well ahead of the 
reference wave. This is similar to the results shown in 
Figure~4 of \inlinecite{cam08}.

\begin{figure}
\centerline{\includegraphics[width=1.0\textwidth]{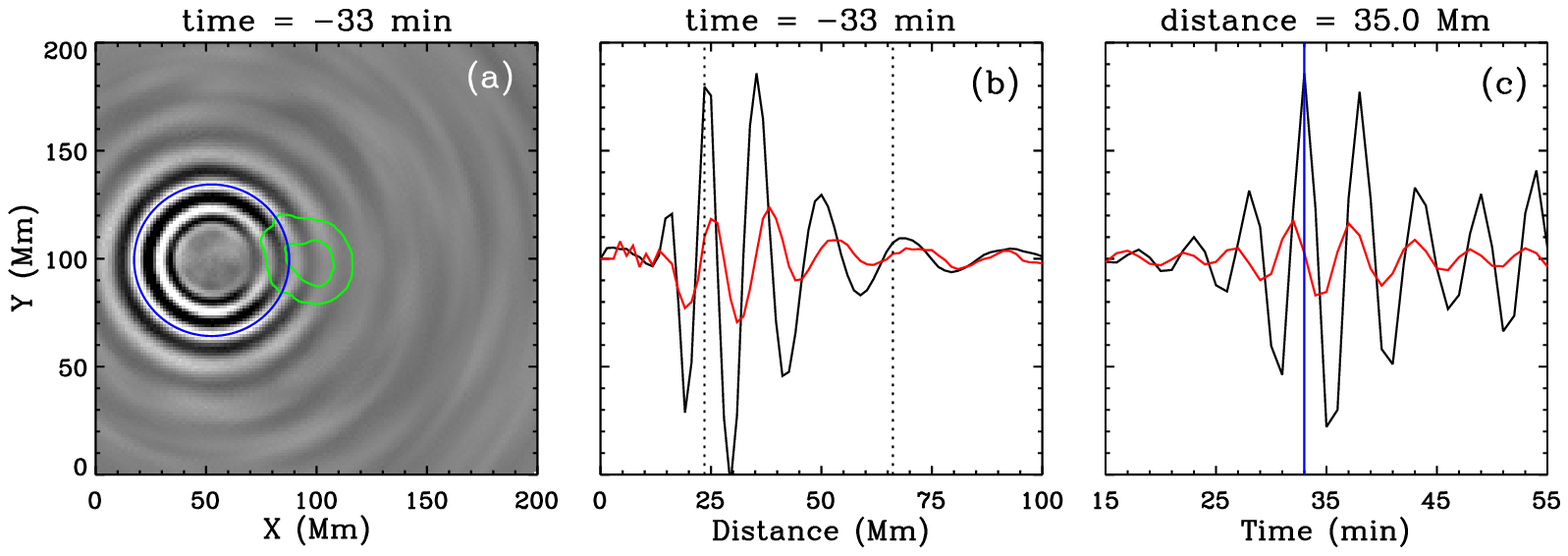} }
\caption{Same as the middle row in Figure~\ref{p_1}, but obtained from the 
negative time lag of the cross-correlation functions. }
\label{p_1_neg}
\end{figure}

Figure~\ref{f_2} shows the results when the wave source is located inside
the sunspot umbra. The selected snapshots display different propagation
times when the entire wave is inside the sunspot boundary, when the wave just 
leaves the sunspot completely, and when the wave is far from 
the sunspot. All of these cases show substantial wave amplitude reductions,
but this may be due to the suppression of oscillations and reduced
wave excitation \cite{par07, cho09}. For all of these cases, despite the 
locations of the wave relative to the sunspot, the first wave peak and 
sometimes the second wave peak are well ahead of the reference wave, 
meaning that the $f$-mode wave originating inside the sunspot umbra always
propagates faster than the waves originating in the quiet Sun.

\begin{figure}
\centerline{\includegraphics[width=0.78\textwidth]{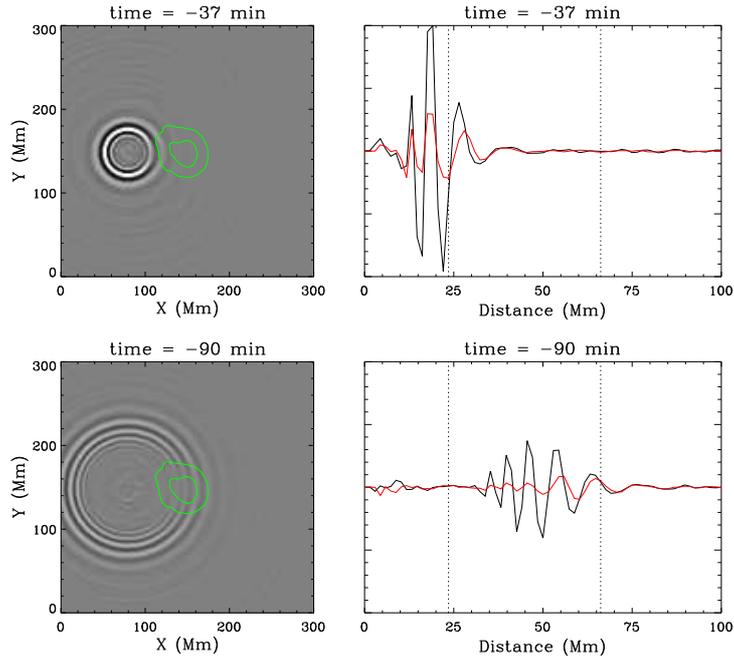} }
\caption{Same as the upper two rows in Figure~\ref{f_1}, but obtained from
the negative lag of the cross-correlation functions. }
\label{f_1_neg}
\end{figure}

\section{Properties of Cross-Correlations with Negative Time Lag}
  \label{S4}
The cross-correlations computed from observed data can have both positive 
and negative time lags. While the positive time lag of the cross-correlations
correspond to waves expanding from the wave source, the negative lag can 
be understood as waves originating from locations surrounding a selected 
wave source and converging toward that source. It is interesting to compare the
results obtained from these negative time lags with those obtained from 
the positive lags.

Figure~\ref{p_1_neg} shows one example of $p$-mode waves when the wave source
is located outside the sunspot. Comparing this figure with the results
for the expanding waves (positive time lag), shown in Figure~\ref{p_1}(d) -- 
(f), one can identify obvious differences between these two sets of results. 
Figure~\ref{p_1_neg}(a) shows that the waveform exhibits an obvious protruding 
shape inside the sunspot, meaning that the waves starting inside
the sunspot and traveling to the outside are significantly faster than the
waves starting from the quiet Sun and traveling into the sunspot.
Figure~\ref{p_1_neg}(b) and (c) more clearly demonstrate this, and the
travel-time difference in this case can be as large as 1.0 minute.

Our results for the converging $f$-mode waves also demonstrate obvious 
differences from the expanding waves, which can be easily seen by 
comparing Figure~\ref{f_1_neg} with Figure~\ref{f_1}. For instance, 
at $\tau = -37$ minute, the wave amplitude is significantly reduced even 
for the waveform outside the sunspot. It shows an obviously faster 
propagation speed inside the sunspot. At $\tau = -90$ minute, the two peaks 
on the right hand side show clear shifts. These are not found, or not clear, 
in the expanding waves. These sharp differences, or asymmetry, 
between the expanding and converging waves, are very interesting and worth 
more study. Of course, the converging waves can be reproduced only in 
numerical simulations with multiple sources.

\section{Summary and Discussion} 
  \label{S5}
We have analyzed the interactions of helioseismic waves with a round sunspot
after separating the $p$- and $f$-mode oscillation signals but without 
using any other signal filters, such as phase-speed filter or ridge filter.
By calculating cross-correlation functions from the observed stochastic 
wavefield, we have reconstructed the wave signals corresponding to effective
point sources (Green's function). For both types of waves, we have studied 
different cases when the wave source is located at different places 
relative to the sunspot.

We summarize our findings as follows:
\begin{itemize}
\item For $p$-mode waves:
\begin{enumerate}
\item When the wave source is located far outside the sunspot, the 
wave passes through the sunspot area with a reduced amplitude and slightly
shorter travel times. Once the wave reaches the other side of the sunspot,
the wave shows little difference from the reference quiet-Sun wave.
However, the negative lags (corresponding to converging waves) of 
cross-correlation functions show a protruding shape inside the sunspot,
meaning that waves originating inside the sunspot propagate faster
toward the outside.
\item When the wave source is located at the boundary of the sunspot,
the wave propagates faster inside the sunspot than in the quiet 
Sun. However, once the wave reaches the other side of the sunspot, 
the waveform becomes very similar to the reference wave as if the wave
does not much effected by the sunspot interior.
\item When the wave source is located inside the sunspot umbra, its travel
time to areas outside of the sunspot is $~\approx 1.0$ minute shorter than 
for the reference waves. The wave amplitude remains substantially reduced 
at all different locations, but this may due to the reduced acoustic 
emissivity inside the active region.
\end{enumerate}
\item For $f$-mode waves: 
\begin{enumerate}
\item When the wave source is outside the sunspot, the wave amplitude
gradually decreases with the propagation through the sunspot. 
The decreased wave amplitude does not recover after the wave reaches
the other side of the sunspot. Before the wave touches the sunspot, and when 
it propagates inside the sunspot, the propagation speed is about the same 
as in the quiet Sun. However, when the wavefront passes the 
sunspot, the part of the wave outside of the sunspot is ahead of the 
reference wave, but the part of the wave still inside the sunspot is behind 
the reference wave. The wavefront remains ahead of the reference
after the wave is completely out of the sunspot area. In contrast, 
the negative lags of cross-correlation functions show faster 
propagation speed when the wave starts from inside the sunspot.
\item If the wave source is located inside the sunspot umbra, the $f$-mode 
wave stays ahead of the reference wave even when the wave travels 
very far outside the sunspot, but this is only true for the leading
part of the waveform. The trailing part of the waveform is often 
in accord with the reference wave. 
\end{enumerate}
\end{itemize}

This study presents a clear picture of how helioseismic waves, both $f$-
and $p$-modes, interact with a sunspot. These results will be useful
for comparing MHD simulations of various sunspot models (e.g.
\opencite{kho09}; \opencite{par09}; \opencite{cam10}). However, more 
importantly, these
pictures also pose some challenges to our current understanding of the 
observed travel-time shifts in the sunspot area.

The $f$-mode results are particularly puzzling. While the faster propagation 
of the $f$-mode waves from the sunspot to the outside can be explained by 
sunspot outflows or moat flows around the sunspot area \cite{giz00}, it is 
difficult to explain the results when the wave source is located outside
of the sunspot. As shown in Figure~\ref{f_1}, no slowness of $f$-mode waves
is observed before the wave encounters the sunspot and during the wave's 
propagation inside the sunspot. The faster propagation is seen only 
in the leading part of the waveform, and starts only when the wave 
appears on the other side of the sunspot. This seems to be inconsistent
with the effect of symmetrical outflows.

The apparent travel-time asymmetry between the $p$-mode waves propagating 
into and out from the sunspot can be clearly seen in Figures~\ref{p_1}
and \ref{p_1_neg}, but this asymmetry is not seen when waves are 
outside of the sunspot area. The travel-time asymmetry inside the sunspot area
has been widely studied and discussed. It was interpreted by a combination 
of subsurface wave-speed perturbations and flow fields \cite{kos00, zha10},
and by a magnetic ``shower-glass" effect \cite{lin05}. However, for a 
quantitative interpretation it requires a careful comparison of the observed 
waveform variations with results of numerical simulations for various sunspot 
models.  For consistency with the observations such simulations should be 
performed for randomly distributed sources, taking into account the non-uniform
distribution of sources and changes in their spectral characteristics
in strong magnetic field regions. This is important because the non-uniform 
distribution of sources may affect the cross-correlation signals \cite{raj06}.

In addition to such a forward modeling approach, our results show perspectives
for developing a new local helioseismology technique: waveform heliotomography.
This technique is similar to the seismic waveform tomography in geophysics.
Instead of inverting travel times, the waveform tomography inverts both
the amplitude and phase information of point-source or cross-correlation 
signals. It has not been used in seismology as much as the travel-time 
tomography because it is computationally intensive. However, with faster 
computers this method becomes more achievable. Our results show that 
the variations of the waveform, in particular the waveform amplitude, 
are significantly larger than the variations of the travel times. This 
means that the wavefront amplitude is probably more sensitive to the sunspot
parameters, and, therefore, may provide more information about the sunspot 
structure and flows. On the other hand, the strong waveform variations 
may require non-linear inversion techniques \cite{sch09}. We plan to 
explore the waveform heliotomography approach in future work.

\end{article}
\end{document}